\documentstyle[aps,version2,preprint]{revtex}    
\begin{document}

\draft

\begin{title}
New perturbation theory of low-dimensional quantum liquids II:\\
operator description of Virasoro algebras in integrable systems
\end{title}

\author{J. M. P. Carmelo $^{1,2,*}$, A. H. Castro Neto $^{2}$,
and D. K. Campbell $^{2}$}
\begin{instit}
$^{1}$ Instituto de Ciencia de Materiales, C.S.I.C.,
Cantoblanco, SP - 28949 Madrid, Spain \\
$^{2}$ Department of Physics, University of Illinois at
Urbana -- Champaign, \\
1110 West Green Street, Urbana, Illinois 61801-3080
\end{instit}
\begin{instit}
\end{instit}
\receipt{February 1994}

\begin{abstract}
We show that the recently developed {\it pseudoparticle operator
algebra} which generates the low-energy Hamiltonian eigenstates
of multicomponent integrable systems also provides a natural
operator representation for the the Virasoro algebras associated
with the conformal-invariant character of the low-energy spectrum
of the these models. Studying explicitly the Hubbard chain in a
non-zero chemical potential and external magnetic field, we
establish that the pseudoparticle perturbation theory provides a
correct starting point for the construction of a suitable
critical-point Hamiltonian. We derive explicit expressions
in terms of pseudoparticle operators for the generators of the
Virasoro algebras and the energy-momentum tensor, describe the
conformal-invariant character of the critical point from the
point of view of the response to curvature of the two-dimensional
space-time, and discuss the relation to Kac-Moody algebras
and dynamical separation.
\end{abstract}
\renewcommand{\baselinestretch}{1.656}   

\pacs{PACS numbers: 64.60. Fr, 03.65. Nk, 05.70. Jk, 72.15. Nj}

\narrowtext
\section{INTRODUCTION}

In this paper we continue our study of a new perturbation
theory for low-dimensional quantum liquids by showing
that the pseudoparticle operator algebra and
perturbation theory introduced in the preceding paper (Ref.
\cite{Carmelo94}, henceforth I) leads, in a natural way, to the
correct operator description of the critical-point physics of
multicomponent integrable quantum liquids
\cite{Yang,Lieb,Izergin,Frahm,Neto}.

Here by ``critical point''
we mean that regime in which conformal invariance holds
in the continuum limit.
Among the specific physical quantities of interest at
the critical point are the velocities of the gapless modes,
matrix elements of various currents coupled to experimental
probes, and correlation functions for charge, spin, and
other observable operators.

Considerable previous work has established that conformal invariance
provides particularly powerful insight into massless field theories
in two space-time dimensions \cite{Belavin,Blote,Boy,Houches} and
hence into the critical-point physics of integrable quantum
liquids.
In particular, at the critical point these theories are characterized
by the ``conformal anomaly,'' $c$,
which can be extracted from either \cite{Blote,Houches}

(i) The finite-size-scaling behavior of the ground-state energy; or

(ii) The stress tensor-tensor correlation function associated
with the response to curvature of the two-dimensional space.

On the one hand, it was shown from the analysis of
finite-size-scaling results of these systems
\cite{Izergin,Frahm} that each gapless excitation corresponds to one
Virasoro algebra with conformal anomaly $c=1$ and that the complete
critical theory is given as a direct product of the Virasoro
algebras.

On the other hand, method (ii) has {\it not} been
applied to multicomponent models solvable by Bethe ansatz
(BA), since conformal-field theory refers to
Lorentz-invariant systems only. In the multicomponent quantum
liquids there are several different ``light velocities''
and it is not immediately clear how we can take all these velocities to
be simultaneously equal to one. One of our tasks in
this article will be to clarify this particular point.

For definiteness, as in I we focus our discussion on the
Hubbard chain in a non-zero chemical potential and an external magnetic
field, which corresponds to a two component quantum liquid.
Further, again as in I, we limit our study to the $U(1)\otimes U(1)$ sector of
parameter space where the low-energy physics is dominated
by lowest-weight states (LWS) of both the $\eta$ spin and
spin algebras \cite{Korepin}. Although our
explicit calculations are for the Hubbard  model, our general
results and approach apply to multicomponent integrable
systems \cite{Yang,Lieb,Izergin,Frahm,Neto}.

The critical-point energy spectrum of the Hubbard chain
in a magnetic field has been obtained previously by combining
the BA solution, finite-size studies, and
conformal-field theory \cite{Frahm,Woy}. Further, the asymptotic
behavior of correlation functions and the corresponding critical
exponents in the case of multicomponent-integrable
quantum liquids have also been obtained \cite{Frahm}.

Our goal in this article is to extend these results  -- which have
dealt with energies and expectation values only -- to
a full operator representation of the critical-point
theory. In particular, we will show that the perturbative character
of the pseudoparticle basis studied in I permits
full, explicit evaluation of the properties of the critical
point and in particular allows the {\it operator
description} of the Virasoro algebras of integrable quantum
liquids. This permits us to construct, and to calculate using
simple operator forms, the stress tensor and generators of
the Virasoro algebras \cite{Neto} and to characterize the
highest-weight states (HWS) \cite{Neto,Blote,Boy,Houches} of
these algebras; such states correspond to the ``primary fields''
\cite{Boy,Houches}.

Before outlining our approach in this paper, let us recall
briefly the crucial results from I. As in the Landau
Fermi-liquid approach, we use the interacting
ground state as the reference state \cite{Pines,Baym}.
Importantly, in the pseudoparticle basis, this reference
state has the form of a simple Slater determinant and
further the pseudoparticle interaction $f$ functions
are, in general, non-singular.
This allows the introduction of a well-defined pseudoparticle
perturbation theory. Since the one-particle spectral
function is fully incoherent -- {\it i.e.}, $Z_F = 0$ --
the system is {\it not} a Fermi liquid but has what we
have termed a ``Landau-liquid'' character
\cite{Carmelo91,Carmelo91b,Carmelo92,Carmelo92b,Carmelo92c,Campbell}.

In the pseudoparticle basis, the low-energy physics is fully
determined by two-pseudoparticle (zero-momentum) forward
scattering.
Near the pseudo-Fermi points we can classify the few types
of pseudoparticle scattering processes which are left over.
In order to describe the critical-theory spectrum we
construct an effective
Hamiltonian containing {\it only}
the relevant two-pseudoparticle scattering terms of the full
Hamiltonian (see Eq. $(1)$ below).  This is done by
linearizing the {\it pseudoparticle bands} of the Hamiltonian
in the pseudoparticle basis and considering only the values of
the two-pseudoparticle interaction at the pseudo-Fermi points.
Hence, the critical-point Hamiltonian is constructed directly
{\it in the pseudoparticle basis}, in contrast to the usual
procedure of deriving the critical-point Hamiltonian in the
electronic basis and normal-ordered relative to the electronic
non-interacting ground state \cite{Castro}.

The remainder of the paper is organized into five
additional sections. In Sec. II we construct
the critical-point Hamiltonian and show that the
second-order pseudoparticle perturbation theory
leads to the energy spectrum of conformal-field theory.
In Sec. III we describe the conformal-invariant character of
the  multicomponent integrable quantum liquids from the
point of view of the response to the curvature of the
two-dimensional space-time. Further, we use
the pseudoparticle operator basis to show that at the critical
point the energy-momentum tensor operator decouples into
two (or $\nu$, for the general multicomponent case)
new tensors which act on orthogonal
Hilbert subspaces. Each gapless excitation branch corresponds
to an independent Minkowski space (each with common space and
time but a different ``light'' velocity). The
Lorentz-invariance in each of these spaces is associated
with independent Virasoro algebras and two (or $\nu$)
related affine-Lie algebras (Kac-Moody algebras)
\cite{Neto,Boy,Houches}. We also introduce the
pseudoparticle field theory, which describes the quantum
liquid at low energy and small momentum.
In Sec. IV we write in operator form the generators of the
Virasoro algebras. The treatment  of the associated affine-Lie
algebras and of the dynamical separation is
presented in Sec. V. Finally, Sec. VI contains the discussion
and concluding remarks.

\section{CRITICAL-POINT EFFECTIVE HAMILTONIAN}

In this section we use the perturbative character of the
pseudoparticle basis to derive a critical-point
Hamiltonian. This is constructed from the
Hamiltonian in the pseudoparticle basis normal-ordered
with respect to the ground state of the many-particle
problem. As shown in I, the normal-ordered Hamiltonian
has an infinite number of terms and is given by \cite{Carmelo94}

\begin{equation}
:\hat{H}: = \sum_{i=1}^{\infty}\hat{H}^{(i)} \, ,
\end{equation}
where the first- and second-order pseudoparticle scattering
terms read

\begin{equation}
\hat{H}^{(1)} = \sum_{q,\alpha}
\epsilon_{\alpha}(q):\hat{N}_{\alpha}(q): \, ,
\end{equation}
and

\begin{equation}
\hat{H}^{(2)} = {1\over {N_a}}\sum_{q,\alpha} \sum_{q',\alpha'}
{1\over 2}f_{\alpha\alpha'}(q,q')
:\hat{N}_{\alpha}(q)::\hat{N}_{\alpha'}(q'): \, ,
\end{equation}
respectively. Here $\epsilon_{\alpha}(q)$ are the $\alpha$
pseudoparticle bands \cite{Carmelo91b,Carmelo92b} and the four
(or, in the general case, $\nu\times\nu$) ``Landau'' $f$
functions have the universal form

\begin{eqnarray}
f_{\alpha\alpha'}(q,q') & = & 2\pi v_{\alpha}(q)
\Phi_{\alpha\alpha'}(q,q')
+ 2\pi v_{\alpha'}(q') \Phi_{\alpha'\alpha}(q',q) \nonumber \\
& + & \sum_{j=\pm 1} \sum_{\alpha'' =c,s}
2\pi v_{\alpha''} \Phi_{\alpha''\alpha}(jq_{F\alpha''},q)
\Phi_{\alpha''\alpha'}(jq_{F\alpha''},q') \, ,
\end{eqnarray}
where the pseudoparticle group velocities are given by

\begin{equation}
v_{\alpha}(q) = {d\epsilon_{\alpha}(q) \over {dq}} \, ;
\hspace{1cm} v_{\alpha}\equiv v_{\alpha}(q_{F\alpha}) \, ,
\end{equation}
and the velocities $v_{\alpha}$ play a determining role at
the critical point, representing the ``light'' velocities
which appear in the conformal-invariant expressions
\cite{Carmelo94,Frahm,Neto,Carmelo91b}. $\Phi_{\alpha\alpha'}(q,q')$
is the two-pseudoparticle forward-scattering phase shift
(see Eqs. $(20)-(23)$ below and Eqs. $(23)-(26)$ of Ref.
\cite{Carmelo92b}). The form of the
normal-ordered Hamiltonian $(1)-(3)$ is {\it universal}
for the class of integrable multicomponent quantum
liquids \cite{Carmelo94}.

The ground state associated with a canonical
ensemble of $(\eta_z,S_z)$ values (here $\eta_z$
and $S_z$ are the eigenvalues of the diagonal
generator of the $\eta$ spin and spin $SU(2)$
algebras, respectively -- see Eqs. $(9)$ and $(10)$ of
paper I) has the form \cite{Carmelo94}

\begin{equation}
|0;\eta_z,S_z\rangle = \prod_{\alpha=c,s}
[\prod_{q=q_{F\alpha }^{(-)}}^{q_{F\alpha }^{(+)}}
b^{\dag }_{q\alpha }]
|V\rangle \, ,
\end{equation}
where the operator $b^{\dag }_{q\alpha }$ creates one
$\alpha$ pseudoparticle of pseudomomentum $q$,
$|V\rangle$ is the $n=0$ electronic and pseudoparticle
vacuum and $q_{F\alpha }^{(\pm)}$ are the pseudo-Fermi
points. When the number of $\alpha$ pseudoparticles,
$N_{\alpha }$, is odd (even) and the
quantum numbers \cite{Carmelo94} $I_j^{\alpha }$
are integers (half integers) these are symmetric and given by
\cite{Carmelo94}

\begin{equation}
q_{F\alpha }^{(+)}=-q_{F\alpha }^{(-)} =
{\pi\over {N_a}}[N_{\alpha}-1] \, ,
\end{equation}
whereas when $N_{\alpha }$ is odd (even) and
$I_j^{\alpha }$ are half integers (integers)
we have that

\begin{equation}
q_{F\alpha }^{(+)} = {\pi\over {N_a}}N_{\alpha }
\, , \hspace{1cm}
-q_{F\alpha }^{(-)} ={\pi\over {N_a}}[N_{\alpha }-2] \, ,
\end{equation}
or

\begin{equation}
q_{F\alpha }^{(+)} = {\pi\over {N_a}}[N_{\alpha }-2]
\, , \hspace{1cm}
-q_{F\alpha }^{(-)} = {\pi\over {N_a}}N_{\alpha } \, .
\end{equation}
(Similar expressions are obtained for the pseudo-Brioullin
zones limits $q_{\alpha }^{(\pm)}$ if we replace
in Eqs. $(12)-(14)$ $N_{\alpha }$ by $N_{\alpha }^*$.)
In many expressions the quantities $(7)-(9)$
are replaced by the leading order term,
$q_{F\alpha }^{(\pm)}\approx \pm q_{F\alpha }$,
where $q_{F\alpha }$ is given by

\begin{equation}
q_{F\alpha } = {\pi N_{\alpha }\over {N_a}}
\, .
\end{equation}

As discussed in detail in I, in the $U(1)\otimes U(1)$
sector of parameter space and for energy scales
smaller than the gaps of the non-LWS multiplets and
LWS II \cite{Carmelo94,Two,Carmelo94a,Nuno}, the Hilbert
space of the Hamiltonian in $(1)-(3)$ {\it coincides} with
the full Hilbert space of the quantum problem \cite{Carmelo94},
so that it provides a complete description of the low-energy
physics and can be used as starting point for the construction
of a critical-point Hamiltonian.

This construction proceeds by linearizing the
pseudoparticle bands $\epsilon_{\alpha}(q)$ of the above
Hamiltonian around the pseudo-Fermi points and keeping only the
two-pseudoparticle interaction term $(3)$ \cite{Neto}.
In addition, one replaces the full $f$ function expressions $(4)$
by the corresponding values at the pseudo-Fermi points.
As shown previously, the forms of these $f$ functions
imply that at the critical point the two-pseudoparticle
phase shifts play a crucial role. Measuring
the pseudomomentum from the pseudo-Fermi points adds
the index $\iota=sgn (q)1=\pm 1$, which defines
the right ($\iota=1$) and left ($\iota=-1$)
movers, to the pseudoparticle operators, which
also depend on the pseudomomentum $q$ and the color $\alpha$.
The pseudoparticle operators $b^{\dagger}_{q\alpha}$
($b_{q\alpha}$) become $b^{\dagger}_{\kappa\alpha\iota}$
($b_{\kappa\alpha\iota}$) and the pseudoparticle number operators
$\hat{N}_{\alpha}(q)=b^{\dagger}_{q\alpha}b_{q\alpha}$
become

\begin{equation}
\hat{N}_{\alpha,\iota}(\kappa) =
b^{\dagger}_{\kappa\alpha\iota}b_{\kappa\alpha\iota}
\, ,
\end{equation}
where the new pseudomomentum $\kappa$ is such that

\begin{eqnarray}
\kappa & = & q - q_{F\alpha}^{(+)} \, , \hspace{1cm}
-q_{F\alpha}^{(+)}<\kappa <(q_{\alpha}^{(+)} -
q_{F\alpha}^{(+)}) \, , \hspace{1cm} \iota = 1 \, ,
\nonumber\\
& = & q - q_{F\alpha}^{(-)} \, , \hspace{1cm}
(q_{\alpha}^{(-)} - q_{F\alpha}^{(-)})<\kappa <
-q_{F\alpha}^{(-)} \, , \hspace{1cm} \iota = -1 \, ,
\end{eqnarray}
and the pseudo-Fermi points $q_{F\alpha}^{(\pm)}$
are given by Eqs. $(7)-(9)$.
The number of $\alpha,\iota$ pseudoparticle operator
is given by $\hat{N}_{\alpha,\iota} =
\sum_{\kappa}\hat{N}_{\alpha,\iota}(\kappa)$.

In normal order relative to the ground state $(6)$,
the operators $(11)$ are thus given by \cite{Neto}

\begin{equation}
:\hat{N}_{\alpha,\iota}(\kappa ): = \hat{N}_{\alpha,\iota}(\kappa )
- \Theta (-\iota\kappa ) \, .
\end{equation}

The critical-point Hamiltonian reads

\begin{eqnarray}
:\hat{{\cal H}}: & = & \sum_{\kappa,\alpha ,\iota}
\iota\kappa v_{\alpha}:\hat{N}_{\alpha,\iota}(\kappa):
+ {1\over {N_a}}\sum_{\kappa,\kappa'}\sum_{\alpha,\alpha'}
\sum_{\iota}{1\over 2}[ f_{\alpha\alpha'}^{1}
:\hat{N}_{\alpha,\iota}(\kappa )::\hat{N}_{\alpha',\iota}
(\kappa'): \nonumber \\
& + & f_{\alpha\alpha'}^{-1}
:\hat{N}_{\alpha,\iota}(\kappa ):
:\hat{N}_{\alpha',-\iota}(\kappa' ):] \, ,
\end{eqnarray}
where the two (or $\nu$) ``light velocities'' $v_{\alpha}$
are given in Eq. $(5)$
and

\begin{equation}
f_{\alpha\alpha'}^{1} =
f_{\alpha\alpha'}(q_{F\alpha}^{(\pm)},q_{F\alpha'}^{(\pm)})
\, , \hspace{1cm}
f_{\alpha\alpha'}^{-1} =
f_{\alpha\alpha'}(q_{F\alpha}^{(\pm)},q_{F\alpha'}^{(\mp)}) \, ,
\end{equation}
are two values of the $f$ functions.
$f_{\alpha\alpha'}^{1}$ ($f_{\alpha\alpha'}^{-1}$) refers
to forward-scattering of two right-moving or two left-moving
pseudoparticles (one right-moving and one left-moving
pseudoparticles) close to the corresponding pseudo-Fermi
points.

To clarify the scale-invariant character of the
critical-point Hamiltonian $(14)$ it is useful to express the $f$
functions $f_{\alpha\alpha'}^{\pm 1}$ $(15)$ as

\begin{equation}
f_{\alpha\alpha'}^{\pm 1} = 2\pi\sum_{\alpha''}2v_{\alpha''}
G_{\alpha''}^{\pm 1}(\alpha,\alpha') \, ,
\end{equation}
where the two-pseudoparticle interaction functions
$G_{\alpha }^{1} (\alpha',\alpha'')$ and $G_{\alpha }^{-1}
(\alpha',\alpha'')$ are given explicitly by \cite{Neto}

\begin{equation}
2G_{\alpha }^{1}(\alpha',\alpha'') =
-\delta_{\alpha ,\alpha'}\delta_{\alpha ,\alpha''}
+ {1\over 2}[\xi_{\alpha \alpha '}^0\xi_{\alpha \alpha ''}^0
+ \xi_{\alpha \alpha '}^1\xi_{\alpha \alpha ''}^1] \, ,
\end{equation}
and

\begin{equation}
2G_{\alpha }^{-1}(\alpha',\alpha'') =
{1\over 2}[\xi_{\alpha \alpha '}^0\xi_{\alpha \alpha ''}^0
- \xi_{\alpha \alpha '}^1\xi_{\alpha \alpha ''}^1] \, ,
\end{equation}
respectively, and the dimensionless parameters $\xi_{\alpha
\alpha '}^j$ (with $j=0,1$) are simple combinations of the
two-pseudoparticle phase shifts at the pseudo-Fermi
points

\begin{equation}
\xi_{\alpha \alpha ' }^j = \delta_{\alpha \alpha '} +
\Phi_{\alpha \alpha '}(q_{F\alpha}, q_{F\alpha '}) +
(-1)^j\Phi_{\alpha \alpha '}(q_{F\alpha},
-q_{F\alpha '}) \, ,
\end{equation}
where the phase shifts $\Phi_{\alpha \alpha '}(q,q')$ are
given by

\begin{equation}
\Phi_{cc}(q,q')=
{\bar{\Phi}}_{cc}({\sin K_0(q)\over u},{\sin K_0(q')\over u})
\, ,
\end{equation}

\begin{equation}
\Phi_{cs}(q,q')=
{\bar{\Phi}}_{cs}({\sin K_0(q)\over u},S_0(q'))
\, ,
\end{equation}

\begin{equation}
\Phi_{sc}(q,q')=
{\bar{\Phi}}_{sc}(S_0(q),{\sin K_0(q')\over u})
\, ,
\end{equation}
and

\begin{equation}
\Phi_{ss}(q,q')=
{\bar{\Phi}}_{ss}(S_0(q),S_0(q'))
\, ,
\end{equation}
and $K_0(q)$ and $S_0(q)$ are the ground-state
solution of Eqs. (A1) and (A2) of paper I. The
phase shifts ${\bar{\Phi}}_{cc}(x,x')$,
${\bar{\Phi}}_{cs}(x,y')$, ${\bar{\Phi}}_{sc}(y,x')$,
and ${\bar{\Phi}}_{ss}(y,y')$ are the solutions
of the four coupled integral equations $(32)-(35)$
of Ref. \cite{Carmelo92b}. For a study of the relation
of these phase shifts to two-electron matrix
elements see Ref. \cite{Carmelo92c}.

Inserting $(18)$ into the RHS of Eq. $(14)$ leads to
\begin{eqnarray}
:\hat{{\cal H}}: & = & \sum_{\alpha ,\iota=\pm 1}v_{\alpha}\{\iota
\sum_{\kappa}\kappa  :\hat{N}_{\alpha,\iota}(\kappa):
\nonumber \\
& + & {2\pi\over {N_a}}\sum_{\kappa,\kappa'}\sum_{\alpha',\alpha''}
[G_{\alpha }^{1}(\alpha',\alpha'')
:\hat{N}_{\alpha',\iota}(\kappa )::\hat{N}_{\alpha'',\iota}
(\kappa'):\nonumber \\
& + & G_{\alpha }^{-1}(\alpha',\alpha'')
:\hat{N}_{\alpha',\iota}(\kappa )::\hat{N}_{\alpha'',-\iota}
(\kappa'):]\}\, ,
\end{eqnarray}
which is precisely the critical-point Hamiltonian presented in our
earlier work \cite{Neto}. It includes two-pseudoparticle (momentum
$k=0$) forward-scattering only.

Note that the Hamiltonian $(24)$ can be written
as $:\hat{{\cal H}}:=\sum_{\alpha} :\hat{{\cal H}}^{\alpha}:$.
Remarkably, despite the interaction terms, each term
$:\hat{{\cal H}}^{\alpha}:$ in the Hamiltonian is {\it scale
invariant}, the only scale being the ``light velocity'' $v_{\alpha
}$. In the renormalization group terms, this means
that we are working {\it at the interacting fixed point}
\cite{Neto}. Since we have considered the relevant
two-pseudoparticle interactions only, it follows from the
pseudoparticle perturbation theory presented in
I that the critical-point spectrum
corresponds, {\it exclusively}, to the second-order energy
expansion in the density of excited pseudoparticles.

At {\it small} momentum and low energy the only relevant
term in the Hamiltonian $(14),(24)$
is the non-interacting pseudoparticle term. This explains the
Landau-liquid character of the problem in the pseudoparticle
basis. In contrast, in the case of excitations with {\it large}
momentum and low energy or excitations involving
small changes in the numbers of $\alpha$ type
pseudoparticles both the non-interacting and the
two-pseudoparticle terms of $(14),(24)$ must be included.

{}From the adiabatic continuity principle studied in Refs.
\cite{Carmelo92,Carmelo92c}, which refers to
small momentum and low energy excitations only, it follows that,
in the limit of zero onsite electronic interaction $U$,
the non-interacting term of the critical point Hamiltonian
becomes the corresponding non-interacting critical-point Hamiltonian
{\it in terms of electron operators}.
Further, in this limit the colors $\alpha$ become
the spin projections $\sigma (\alpha)$ and the
velocities $v_{\alpha}$ the Fermi velocities
$v_{F\sigma (\alpha)}$ \cite{Carmelo92c}.
Note, however, that this does not hold true when we
consider low-energy excitations of large momentum
or involving changes in the numbers of
pseudoparticles. In this case the interaction
term of the Hamiltonian $(14),(24)$ becomes relevant.
Importantly, in the present model and in
multicomponent systems, that pseudoparticle interaction
term does not necessarily vanish in the $U\rightarrow 0$
limit \cite{Carmelo94}. This is related to
the exotic properties of the electron - pseudoparticle
canonical transformation \cite{Two,Carmelo94a}.

At the critical-point, the ground-state fluctuations of the finite
system correspond, in the infinite system, to low-energy
excitations. In the case of the fluctuations associated with
the response to curvature of the space-time, which are related
to Lorentz invariance, the relevant critical-point
excitations have small momentum and low energy. This
corresponds to the usual definition of the critical
point.

On the other hand, however, the scale-invariant character
of the Hamiltonian $(24)$ suggests that considering {\it all}
the low-energy excitations of the infinite system,
including eigenstates with large momentum and/or with
different numbers $N_{\alpha}$ of $\alpha$ pseudoparticles,
should lead to the same energy spectrum as the finite-size
studies of the finite system \cite{Frahm}. Our second-order
pseudoparticle perturbation theory, which is equivalent to the
use of the Hamiltonian $(14),(24)$, confirms this suggestion.

Therefore, we can define the critical point in a broader
way by considering the low-energy Hilbert space associated
with the conformal-invariant character of the quantum liquid
\cite{Frahm,Neto}. In this more general case, we need to consider
the Hilbert space spanned by {\it all} Hamiltonian eigenstates of
low energy. This Hilbert space can be divided into three
Hilbert subspaces which we call (A), (B), and (C) \cite{Neto}.
The corresponding BA Hamiltonian eigenstates were
used in Refs. \cite{Izergin,Frahm,Woy} to derive the conformal spectrum
of the Hubbard chain. Here we intend to
characterize these excitations in terms of the operators
which generate them from the ground state
$(6)$, which in the pseudoparticle basis
can be rewritten in terms of right and left $\alpha$
pseudoparticles as

\begin{equation}
|0;\eta_z,S_z\rangle = \prod_{\alpha ,\iota}
[\prod_{-\iota\kappa>0}
b_{\kappa\alpha\iota}^{\dagger}]|V\rangle \, .
\end{equation}

The three types of eigenstates that span the entire
low-energy Hilbert space include:

(A) Eigenstates associated with small changes $\delta N_{\alpha}$
in the numbers of $\alpha$ pseudoparticles $N_{\alpha}$.
These states correspond to an ``adiabatic'' change
from a given point of the $U(1)\otimes U(1)$ sector of parameter
space to a neighboring point of the same space. Since these
excitations involve changes in the values of the numbers
$N_{\alpha}$, they imply a change in the values of the pseudo-Fermi points
$q_{F\alpha}^{(\pm)}$ $(7)-(9)$. This can change the integer
or half-integer character of the quantum numbers $I_j^{\alpha}$
which, as discussed in detail in I, determine the nature of the
BA eigenstate.
We assume that the density of pseudoparticles ``added'' or ``removed'',
$\delta n_{\alpha}={\delta N_{\alpha}\over {N_a}}$, is small.
Note that the number of $\alpha$ pseudoparticles is
a good quantum number. In addition, what is added,
removed, or spin-flipped are electrons, not pseudoparticles
\cite{Carmelo94}.

These low-energy states in the Hilbert space are in fact {\it ground
states} of the form $(25)$ of neighboring parameter-space
points. The density $\delta n_{\alpha}$ can be expressed in
terms of the changes in the pseudo-Fermi points as
follows $2\pi\delta n_{\alpha}=\delta q_{F\alpha}^{(+)}
-\delta q_{F\alpha}^{(-)}$ \cite{Carmelo94}. When
$\delta n_{\alpha}>0$ these states are of the form

\begin{equation}
|(A)\rangle = \prod_{\alpha}
[\prod_{\kappa=0}^{\delta q_{F\alpha}^{(+)}}
b_{\kappa\alpha 1}^{\dagger}]
[\prod_{\kappa=0}^{\delta q_{F\alpha}^{(-)}}
b_{\kappa\alpha -1}^{\dagger}]
|0;\eta_z,S_z\rangle \, ,
\hspace{1cm} \delta n_{\alpha} > 0 \, ,
\end{equation}
whereas for $\delta n_{\alpha}<0$ they are given by

\begin{equation}
|(A)\rangle = \prod_{\alpha}
[\prod_{\kappa=0}^{\delta q_{F\alpha}^{(+)}}
b_{\kappa\alpha 1}]
[\prod_{\kappa=0}^{\delta q_{F\alpha}^{(-)}}
b_{\kappa\alpha -1}]
|0;\eta_z,S_z\rangle \, ,
\hspace{1cm} \delta n_{\alpha} < 0 \, .
\end{equation}

(B) One-pair and multipair eigenstates involving
the transfer of a small density of $\alpha$- pseudoparticles
from the pseudo-Fermi points $q=q_{F\alpha}^{(\pm)}$
to $q=q_{F\alpha}^{(\mp)}\mp {2\pi\over {N_a}}$. We introduce
the number $D_{\alpha}$, such that $2D_{\alpha}=\delta N_{\alpha}^-
=[\delta N_{\alpha ,\iota=1}-\delta N_{\alpha ,\iota=-1}]$,
which is related to the number of $\alpha$- pseudoparticles
transferred. These eigenstates have large momentum and ``transform''
$\alpha$, $\iota$ pseudoparticles into $\alpha$, $-\iota$
pseudoparticles. When $D_{\alpha}>0$ and for small but
finite densities of pseudoparticles transferred these states can
be generated from the ground state $(25)$ as

\begin{equation}
|(B)\rangle = \prod_{\alpha}
[\prod_{\kappa=0}^{2\pi D_{\alpha}/N_a}
b_{\kappa\alpha 1}^{\dagger}
b_{\kappa\alpha -1}]|0;\eta_z,S_z\rangle \, ,
\hspace{1cm} D_{\alpha} > 0 \, ,
\end{equation}
whereas if $D_{\alpha}<0$ and for small finite
densities
they are of the form

\begin{equation}
|(B)\rangle = \prod_{\alpha}
[\prod_{\kappa=0}^{2\pi D_{\alpha}/N_a}
b_{\kappa\alpha -1}^{\dagger}
b_{\kappa\alpha 1}]|0;\eta_z,S_z\rangle \, ,
\hspace{1cm} D_{\alpha} < 0 \, .
\end{equation}
The states (B) have large momentum

\begin{equation}
k = \sum_{\alpha} D_{\alpha}[q_{F\alpha}^{(+)}
+ {2\pi\over {N_a}} - q_{F\alpha}^{(-)}]
= \sum_{\alpha} D_{\alpha} 2q_{F\alpha}
\, ,
\end{equation}
where $q_{F\alpha}$ is given in Eq. $(10)$.

(C) One-pair and multipair eigenstates involving $\alpha$
pseudoparticle-pseudohole processes around the same point
of the pseudo-Fermi surface $q_{F\alpha}^{(\pm)}$,
the numbers of $\alpha$, $\iota$ pseudoparticle-pseudohole
processes of this kind being denoted by

\begin{equation}
N^{\alpha\iota}_{ph} = \iota {N_a\over {2\pi}}[\sum_{p}
\kappa_{p,\alpha,\iota} - \sum_{h} \kappa_{h,\alpha,\iota}] \, ,
\end{equation}
and the density of excited pseudoparticles ${N^{\alpha\iota}_{ph}
\over {N_a}}$ being small. Here $\kappa_{p,\alpha,\iota}$
($\kappa_{h,\alpha,\iota}$) defines the pseudomomentum values of
the pseudoparticles (pseudoholes). These states are of the form

\begin{equation}
|(C)\rangle = L_{-N^{\alpha\iota}_{ph}}^{\alpha ,\iota}
|0;\eta_z,S_z\rangle \, ,
\end{equation}
where the explicit form of the generator
$L_{-N^{\alpha\iota}_{ph}}^{\alpha ,\iota}$
will be discussed below (see Sec. IV, Eq. $(56)$); as the notation
suggests, we shall in fact see that it is related to
generators of the Virasoro algebra associated with the theory.

The eigenstates (A) and (B) are pseudoparticle
collective excitations which include pseudo-Fermi points
displacements. While the states (A) connect two ground states
of neighboring canonical ensembles because the displacements
of the two kinds of $\iota$ pseudoparticles have opposite sign,
the displacements associated with the excitations (B) have
the same sign and lead to the finite excitation momentum $(30)$.
In addition, when in the case of the eigenstates (A) the quantum
numbers $I_j^{\alpha}$
change from integers (half integers) to half integers (integers), these
excitations involve a global pseudomomentum shift which affects
all pseudoparticles of the corresponding $\alpha$ branch.
This leads to a so-called orthogonal catastrophe \cite{Ander}
and explains the fully incoherent -- {\it i.e.}, $Z_F=0$ --
nature of
the one-electron spectral function \cite{Carmelo92c}.

The pure zero-momentum forward-scattering character of the
Hamiltonian $(14),(24)$
implies that the numbers of right
and left $\alpha$ pseudoparticles obey independent
conservation laws \cite{Castro}, i.e. the four (or
$2\nu$) numbers $N_{\alpha ,\iota}$ of $\alpha$, $\iota$
pseudoparticles are good quantum numbers. From the
perspective of Lorentz invariance, this means that the $\iota=1$
and $\iota=-1$ problems are independent and the states
$(26),(27)$ and $(28),(29)$ are seen as $\iota$ ``ground
states'' of different canonical ensembles. Therefore, the
fluctuations which determine the Lorentz invariance involve only
the excitations (C), whereas the states (A) and (B) represent
different choices of starting ground states. In the language
of conformal-field theory this means that (A) and (B) are
HWS of the Virasoro algebras, whereas the excitations (C)
correspond to the towers \cite{Boy}. We will discuss
this in detail below. One consequence of this fact is that
instead of the ground state
$(24)$ we can use as starting states for the excitations
$|(C)\rangle $ both states of form $|(A)\rangle$ and
$|(B)\rangle$.

Starting from any point of the $U(1)\otimes U(1)$
sector of parameter space we can arrive to any other point
of the same sector by performing a set of adiabatic excitations
of type (A). On the other hand, the excitations (B) and (C)
span the low-energy Hilbert space associated with each point
of parameter space (i.e. each canonical ensemble). The Hamiltonian
eigenstates (A)-(C) span the critical-point Hilbert space.
These states are both eigenstates of the full pseudoparticle
Hamiltonian $(1)$
and of the critical-point Hamiltonian $(14),(24)$.
Note, however, that the critical-point Hamiltonian leads to the
correct energy spectrum up to second order in the density
of excited pseudoparticles only. This is because it does
not contain the higher-order terms of the full Hamiltonian
$(1)$ -- according to the pseudoparticle perturbation theory
discussed in Sec. IV of paper I the truncated scattering-second-order
Hamiltonian leads to the correct second-order energy
in the density of excited pseudoparticles.

Since the quantum-liquid Hamiltonian can be expressed exclusively
in terms of the pseudomomentum distribution operators $(11)$
and $(13)$, we can calculate the energy spectrum from the
eigenvalue equations of these operators for the states
$|(A)\rangle$, $|(B)\rangle$, and $|(C)\rangle$ defined by
Eqs. $(26)$, $(27)$, $(28)$, $(29)$, and $(32)$. These
equations are of the form

\begin{eqnarray}
\hat{N}_{\alpha,\iota}(\kappa )|(A)\rangle & = &
\Theta (\delta q_{F\alpha}^{(+)} - \kappa )|(A)\rangle
\hspace{1cm} \iota = 1 \nonumber\\
& = & \Theta (-\delta q_{F\alpha}^{(-)} + \kappa )|(A)\rangle
\hspace{1cm} \iota = -1 \, ,
\end{eqnarray}

\begin{eqnarray}
\hat{N}_{\alpha,\iota}(\kappa )|(B)\rangle & = &
\Theta ({2\pi \over {N_a}}D_{\alpha} - \kappa )|(B)\rangle
\hspace{1cm} \iota = 1 \nonumber\\
& = & \Theta (- {2\pi \over {N_a}}D_{\alpha} + \kappa )|(B)\rangle
\hspace{1cm} \iota = -1 \, ,
\end{eqnarray}
and

\begin{equation}
\hat{N}_{\alpha,\iota}(\kappa )|(C)\rangle =
[\Theta (-\iota\kappa )+{2\pi\over {N_a}}(\sum_{p}\delta
(\kappa -\kappa_{p,\alpha,\iota}) -\sum_{h}\delta (\kappa -
\kappa_{h,\alpha,\iota}))]|(C)\rangle \, .
\end{equation}

In order to calculate the critical-point energy spectrum we
consider an eigenstate $|\Psi\rangle $ involving all three
types of excitations (A), (B), and (C). The corresponding
eigenvalue equation is

\begin{eqnarray}
\hat{N}_{\alpha,\iota}(\kappa )|\Psi\rangle & = &
[\Theta (\delta q_{F\alpha}^{(+)} + {2\pi \over {N_a}}D_{\alpha}
- \kappa ) + {2\pi\over {N_a}}(\sum_{p}\delta
(\kappa -\kappa_{p,\alpha,1}) -\sum_{h}\delta (\kappa -
\kappa_{h,\alpha,1}))]|\Psi\rangle
\hspace{1cm} \iota = 1 \nonumber\\
& = & [\Theta (-\delta q_{F\alpha}^{(-)} -
{2\pi \over {N_a}}D_{\alpha} + \kappa )
+ {2\pi\over {N_a}}(\sum_{p}\delta
(\kappa -\kappa_{p,\alpha,-1})\nonumber\\
& - & \sum_{h}\delta (\kappa -
\kappa_{h,\alpha,-1}))]|\Psi\rangle
\hspace{1cm} \iota = -1 \, .
\end{eqnarray}
By combining the critical-point-Hamiltonian
expression $(14),(24)$ with the eigenvalue Eq. $(36)$
we can calculate the corresponding excitation energy. Expanding
to second order in the density of excited pseudoparticles we find
in units of $2\pi/N_a$ (and in the thermodynamic limit)
\cite{Neto},

\begin{equation}
\Delta E = \langle :\hat{{\cal H}}: \rangle =
\sum_{\alpha,\iota=\pm 1}
v_{\alpha}[h_{\alpha}^{\iota} + N^{\alpha\iota}_{ph}] \, ,
\end{equation}
where

\begin{equation}
h_{\alpha}^{\iota} \equiv {1\over 2} [\sum_{\alpha
'}\xi_{\alpha \alpha '}^1 D_{\alpha '} + \iota\sum_{\alpha
'}\xi_{\alpha \alpha '}^0 (\delta N_{\alpha '}/2)]^2 \, .
\end{equation}
Here the $2\times 2$ (or $\nu\times\nu$) coefficients
$\xi_{\alpha \alpha '}^j$ $(19)$
are the entries of the dressed-charge matrix ${\bf Z}$
when $j=1$ and the entries of the matrix $[{\bf Z}^T]^{-1}$
when $j=0$. (Our definition of the dressed-charge matrix is the
transpose of that used in Refs. \cite{Frahm}.)

We can now relate the low-energy spectrum we have derived
for our critical point Hamiltonian to the form of the
energy spectrum obtained from finite-size scaling
studies \cite{Frahm}: a direct comparison
shows that our expression $(37)$ is {\it precisely}
of the same form as that derived from finite-size studies,
with $h_{\alpha}^{\iota }$ in $(38)$ being
the conformal dimensions of the primary fields
\cite{Neto,Boy}. The perturbative character of the
pseudoparticle basis then implies that it
should provide the correct energy spectrum
up to second order in the density of excited
pseudoparticles.

We emphasize that the excitation energy $(37)$ of the state
$|\Psi\rangle$ is decoupled into two types of terms: the terms
containing the parameters $h_{\alpha}^{\iota}$ are generated
by the excitations (A) and (B), whereas the terms involving
the numbers $N^{\alpha\iota}_{ph}$ arise from the excitations (C).
This implies that the HWS of the Virasoro algebras
\cite{Neto,Boy,Houches} correspond to the pseudoparticle
eigenstates of types (A) and (B), whereas the low-momentum
excitations (C) correspond to the towers, as we have mentioned
previously. Therefore, only the excitations (A) and (B) refer to
{\it primary fields} \cite{Boy}.

At the critical point, which corresponds to second-order
energy spectrum $(37)$, the multipair
eigenstates of the Hilbert sub-space (C) can be described as
a direct product of one-pair eigenstates,
as in a non-interacting system. This is because at
low energy and small momentum the two-pseudoparticle
terms of the Hamiltonian $(14),(24)$ are {\it irrelevant}
because they give zero when acting onto the
eigenstates (C) $(32)$. It follows that the energy $(37)$
is additive in the numbers $N^{\alpha\iota}_{ph}$, {\it i.e.}, the $\alpha$
branches are fully decoupled in the case of the excitations
belonging to the space (C) \cite{Neto}. On the other hand, the
$\iota=1$ and $\iota=-1$ excitations
also refer to orthogonal Hilbert subspaces. Therefore, (C)
decouples into a set of four (or $2\times\nu$) smaller Hilbert subspaces
corresponding to each of the $\alpha ,\iota$ branches. Due to
the orthogonality of these $\alpha ,\iota$ spaces, we can uniquely
define the projections of the Hamiltonian and momentum operators
in each of them.

\section{LORENZ INVARIANCE, THE RESPONSE TO CURVATURE OF SPACE-TIME,
         AND CONFORMAL ANOMALIES}

The response of a system with fixed values of the conserved
quantum numbers  $N_{\alpha ,\iota}$ to the curvature of space-time
can be determined by the ground-state fluctuations of the
finite system \cite{Blote} which correspond in the infinite
system to the low-momentum, low-energy excitations of type (C),
the restriction to fixed  $N_{\alpha ,\iota}$ eliminating the (A)
and (B) excitations. Importantly, the two-pseudoparticle terms of the
Hamiltonian $(14),(24)$ give zero when acting onto
the states (C), so that for these excitations the critical-point
Hamiltonian reduces to a non-interacting pseudoparticle problem

\begin{equation}
:\hat{{\cal H}}_0 : = \sum_{\kappa,\alpha,\iota}
\iota \kappa v_{\alpha}
:\hat{N}_{\alpha,\iota}(\kappa): \, ,
\end{equation}
which can be written as $:\hat{{\cal H}}_0:=\sum_{\alpha}
:\hat{{\cal H}}_0^{\alpha}:$. This non-interacting
form provides a tremendous simplification, enabling us to
establish using straightforward calculations several important
results concerning Lorenz invariance and conformal
anomalies. Before presenting these calculations, let us
make some remarks concerning the nature of the states (C) in
terms of both pseudoparticles and electrons.

In a later section we shall prove (see Eq. $(56)$ below)
that the excitations (C) are generated by acting on the ground
state with products of one-pair $\alpha$, $\iota$ pseudoparticle
operators. These generators creating the states (C) are
{\it interaction dependent} mixtures of the electronic operators,
and their explicit forms will be studied elsewhere
\cite{Two,Carmelo94a}. This interaction dependence is
reflected in the fact that the colors $\alpha$ associated
with the conserved pseudoparticle quantum numbers $N_{\alpha}$
cannot be identified with usual quantum numbers such as charge, spin,
or spin projection \cite{Two,Carmelo94a}. However, $N_{\alpha}$
is a conserved quantum number. Similarly, that the
non-interacting pseudoparticle Hamiltonian $(39)$ describes
{\it interacting electrons} is indicated by the
$U$ dependence of the velocities $v_{\alpha}$ of Eq. $(5)$
\cite{Carmelo91b}.

The adiabatic principle of Ref. \cite{Carmelo92c}
implies that the limit $U\rightarrow 0$ of the
Hamiltonian $(38)$ describes correctly the non-interacting
{\it electrons} in the sector of lowest symmetry $U(1)\otimes U(1)$, which
corresponds to $U=0$, $0<n<1$, and $n_{\uparrow}>n_{\downarrow}$.
In this case, the colors $c$ and $s$ become
the spin projections $\uparrow$ and $\downarrow$,
respectively  \cite{Carmelo92c,Two,Carmelo94a}.
The properties of that transformation also imply that the
limits $\eta_z\rightarrow 0$ or (and) $S_z\rightarrow 0$ of
the energy spectrum $(37)$ provide the correct values for the
corresponding higher symmetry sectors of parameter space.

The non-interacting character
of the Hamiltonian $(39)$ in the pseudoparticle basis
follows from the symmetries of the problem
and is the result of the Lorentz invariance which holds in each
of the Hilbert subspaces $\alpha$, $\iota$ involving
states of  type (C): we find
below that at the critical point the energy-momentum tensor
decouples in four (or $2\times\nu$) new tensors
which act on orthogonal Hilbert subspaces. Each gapless
excitation branch corresponds to independent Minkowski spaces,
each with common space and time but different ``light''
$\alpha$ velocities. The Lorentz-invariance
in each of the four (or $2\times\nu$) $\alpha,\iota$ subspaces of
(C) excitations determines the complete critical theory of
the quantum liquids, which is simply given by the direct product
of two (or $\nu$) Virasoro algebras
(each including the $\iota=1$ and $\iota=-1$  components),
in full agreement with the finite-size results
of Refs. \cite{Izergin,Frahm}. In connection to these algebras there
are two (or $\nu$) affine-Lie algebras (Kac-Moody algebras)
\cite{Neto,Boy,Houches}.

Turning to the calculational details, we apply to our
pseudoparticle Hamiltonian $(39)$ the standard continuum
limit techniques: in the limit of small momentum (long wavelength)
and low energy we can ignore the discrete character of the lattice and map
the problem into a continuum field theory whose fields
are labeled by the colors $\alpha$.
Let us then introduce the pseudoparticle fields $\psi_{\alpha
\iota}^{\dagger}$ and $\psi_{\alpha\iota}$. The Hamiltonian
density which corresponds to the Hamiltonian $(39)$ reads

\begin{equation}
\hat{{\cal H}} = \sum_{\alpha,\iota}
\hat{{\cal H}}^{\alpha}_{\iota} \, ,
\end{equation}
where the individual terms $\hat{{\cal H}}^{\alpha}_{\iota}$

\begin{equation}
\hat{{\cal H}}^{\alpha}_{\iota} =
- i \iota v_{\alpha} [\psi_{\alpha\iota}^{\dagger}
{\partial\over {\partial x}}\psi_{\alpha \iota}]
\, ,
\end{equation}
refer to orthogonal Hilbert subspaces of (C) and, therefore,
correspond to four (or $2\times\nu$) independent field theories.
The total Lagrangian density is given by

\begin{equation}
\hat{{\cal L}} =
\sum_{\alpha,\iota} \psi_{\alpha\iota}^{\dagger}
i \{{\partial\over {\partial t}} +
\iota v_{\alpha} {\partial\over {\partial x}}
\} \psi_{\alpha\iota}
 \, .
\end{equation}
It is convenient to introduce the light-cone combinations

\begin{equation}
x_{\iota}^{\alpha} = {1\over 2}\{t - \iota {x\over {v_{\alpha}}}\}
\, .
\end{equation}
By construction, the pseudoparticle field associated with
each orthogonal Hilbert space of colors $\alpha$ and $\iota$
is such that

\begin{equation}
\psi_{\alpha\iota} =
\psi_{\alpha\iota}(x_{1}^{\alpha},x_{-1}^{\alpha}) \, .
\end{equation}
Then the Lagrangian density $(42)$ can be written as

\begin{equation}
\hat{{\cal L}} =
\sum_{\alpha,\iota} i \psi_{\alpha\iota}^{\dagger}
\partial_{\iota}^{\alpha} \psi_{\alpha\iota}
\, ,
\end{equation}
where

\begin{equation}
\partial_{\iota}^{\alpha} =
{\partial\over {\partial x_{\iota}^{\alpha}}}
\, .
\end{equation}

The classical equations of motions are

\begin{equation}
\partial_{\iota}^{\alpha} \psi_{\alpha\iota} = 0
\, ,
\end{equation}
which implies that

\begin{equation}
\psi_{\alpha\iota} =
\psi_{\alpha\iota}(x_{-\iota}^{\alpha}) \, .
\end{equation}

{}From Eqs. $(40)-(48)$ it is straightforward to find that
the energy-momentum tensor is given by

\begin{equation}
\hat{T} = \sum_{\alpha,\iota} \hat{T}^{\alpha}_{\iota}
\, ,
\end{equation}
where each component tensor $\hat{T}^{\alpha}_{\iota}$ reads \cite{Boy}

\begin{equation}
\hat{T}^{\alpha}_{\iota} = {i\over 2} :\psi_{\alpha
-\iota}^{\dagger} \partial_{\iota}^{\alpha}
\psi_{\alpha -\iota}: \, ,
\end{equation}
and corresponds to one of the independent field
theories which refer to orthogonal Hilbert subspaces of (C).

{}From the equal-time commutation relation

\begin{equation}
\{\psi_{\alpha\iota}^{\dagger}(x,t),\psi_{\alpha'\iota'}(x',t)\}
= \delta_{\alpha ,\alpha'} \delta_{\iota,\iota'} \delta (x-x')
\, ,
\end{equation}
one finds the two-point correlation functions

\begin{equation}
\langle 0;\eta_z,S_z|\psi_{\alpha\iota}^{\dagger}(x^{\alpha}_{-\iota})
\psi_{\alpha'\iota'}(y^{\alpha'}_{-\iota'})|0;\eta_z,S_z\rangle =
-{i\over {\pi}} {\delta_{\alpha ,\alpha'} \delta_{\iota,\iota'}
\over {(x^{\alpha}_{-\iota} - y^{\alpha}_{-\iota} + i\epsilon )}}
\, .
\end{equation}

Let us evaluate the leading singularity term of
operator product expansion associated with the tensor-tensor
correlation function \cite{Neto}. This leading term measures the response
of the quantum problem to the curvature of the two-dimensional
space-time. In one-component conformal-invariant problems
this term introduces the conformal anomaly $c$ \cite{Blote}.
In the case of multicomponent systems, the problem
has not previously been studied. However, the present pseudoparticle
operator basis makes this problem trivial because
of the remarkable decoupling $(49)$. A straightforward
calculation using Wick's theorem shows that

\begin{equation}
\langle 0;\eta_z,S_z|\hat{T}^{\alpha}_{\iota}(x^{\alpha}_{\iota})
\hat{T}^{\alpha'}_{\iota'}(y^{\alpha'}_{\iota'})|0;\eta_z,S_z\rangle =
c_{\alpha\iota} {\delta_{\alpha ,\alpha'} \delta_{\iota ,\iota '}
\over {[2(x^{\alpha}_{\iota} - y^{\alpha}_{\iota} +
i\epsilon )^4]}} \, , \hspace{2cm} c_{\alpha\iota} = 1 \, ,
\end{equation}
where, following the notation of our previous work \cite{Neto},
we have used units of $1/(4\pi^2)$.

On the other hand, since the $\alpha$, $\iota$ Hilbert sub
spaces are orthogonal, each tensor $(50)$ corresponds to an
independent Minkowski space with ``light'' velocity $v_{\alpha}$.
Therefore, we can in each of these independent spaces consider
$v_{\alpha}=1$ in Eq. $(43)$, which leads to \cite{Neto}

\begin{equation}
\langle 0;\eta_z,S_z|\hat{T}^{\alpha}_{\iota}(x_{\iota})
\hat{T}^{\alpha'}_{\iota'}(y_{\iota '}')|0;\eta_z,S_z\rangle =
c_{\alpha\iota} {\delta_{\alpha ,\alpha'} \delta_{\iota,\iota'}
\over {[2(x_{\iota} - y_{\iota} + i\epsilon )^4]}}
\, , \hspace{2cm} c_{\alpha\iota} = 1 \, ,
\end{equation}
where $x_{\iota}={1\over 2}\{t-\iota x\}$.

This confirms the finite-size result of Refs. \cite{Izergin,Frahm}
that the present multicomponent systems correspond to a direct
sum of conformal field theories, each possessing a central charge
equal to one. In the present basis the study of the operator
algebra of these Virasoro algebras becomes a simple problem,
as we show in the following section.

\section{OPERATOR REPRESENTATION OF THE VIRASORO ALGEBRAS}

Each pair of tensors $\hat{T}^{\alpha}_{\iota=\pm 1}(x_{\iota})$
is associated with one Virasoro algebra and the corresponding set
of generators $L_{j}^{\alpha ,\iota}$ \cite{Neto,Boy,Houches}. The
$\hat{T}^{\alpha}_{\iota}(x_{\iota})$ can be expressed in terms
of these generators as follows

\begin{equation}
\hat{T}^{\alpha }_{\iota}(x_{\iota}) = {2\pi\over {N_a^2}}
\sum_{j} L_{j}^{\alpha ,\iota}e^{-2\pi ijx_{\iota}/N_a} \, .
\end{equation}

The ``tower'' of states associated with the terms
$N^{\alpha\iota}_{ph}$ (recall the definition of this
quantity in $(31)$) in the energy spectrum $(37)$ can be
constructed by the action of the generators of the Virasoro
algebras on the HWS \cite{Neto,Boy,Houches} of these algebras.
On the other hand, we have shown that
these ``tower'' terms are associated with the
multipair eigenstates (C). It follows that the generators
$L_{j}^{\alpha ,\iota}$ of the RHS of Eq. $(55)$ annihilate
($j>0$) and create ($j<0$) $\alpha ,\iota$ - pseudoparticle-pseudohole
pairs (C) \cite{Neto}. Thus, rather than referring to electrons, the
generators of the Virasoro algebra $\alpha ,\iota$ refer to the
$\alpha ,\iota$ - pseudoparticles. For example, for $j<0$
these generators are given by

\begin{equation}
L_{j}^{\alpha ,\iota} = \prod_{\kappa_{p,\alpha,\iota},
\kappa_{h,\alpha,\iota}}
[b^{\dag }_{\kappa_{p,\alpha,\iota}\alpha\iota}
b_{\kappa_{h,\alpha,\iota}\alpha \iota}] \, ,
\end{equation}
where the product contains $|j|=N^{\alpha\iota}_{ph}$ factors
with $|j|>0$ giving the number of pseudoparticle-pseudohole
processes (see Eq. $(31)$) and the pseudomomentum values
$\kappa_{h,\alpha,\iota}$ and $\kappa_{p,\alpha,\iota}$ are
the pseudomomentum values associated with these processes and are
the same as in the RHS of Eq. $(31)$.

The $j=0$ generators $L_0^{\alpha ,\iota}$, which {\it are not}
of the form $(56)$, play a particularly important role because
they define the critical-point Hamiltonian and momentum operator
\cite{Boy}. In the present case they have the
universal form

\begin{eqnarray}
L_0^{\alpha ,\iota} & = & \iota\sum_{\kappa}\kappa
:\hat{N}_{\alpha,\iota}(\kappa):
\nonumber \\
& + & {2\pi\over {N_a}}\sum_{\kappa,\kappa'}
\sum_{\alpha',\alpha'',\iota'}[F_{\alpha,
\iota }^{1}(\alpha',\alpha'',\iota')
:\hat{N}_{\alpha',\iota'}(\kappa ):
:\hat{N}_{\alpha'',\iota'}(\kappa'): \nonumber \\
& + & F_{\alpha, \iota }^{-1}(\alpha',\alpha'',\iota')
:\hat{N}_{\alpha',\iota'}(\kappa ):
:\hat{N}_{\alpha'',-\iota'}(\kappa'):] \, ,
\end{eqnarray}
where

\begin{equation}
F_{\alpha, \iota }^{\pm 1}(\alpha',\alpha'',\iota')
= {1\over 2}G_{\alpha }^{\pm 1}(\alpha',\alpha'')
+ {\iota\iota' \over 4}
\xi_{\alpha\alpha'}^1\xi_{\alpha\alpha''}^0 \, ,
\end{equation}
$\xi_{\alpha\alpha'}^j$ is given in Eq. $(19)$ and
the remaining parameters of the RHS of Eqs. $(57)$ and
$(58)$ are the same as in Eq. $(24)$.
The conformal- and scale-invariant character of each $\alpha$
term of the Hamiltonian $(24)$ allows us to take $v_{\alpha}=1$
in Eq. $(57)$. The critical-point Hamiltonian $(24)$
can be expressed in terms of the generators $(57)$ as

\begin{equation}
:\hat{{\cal H}}: = \sum_{\alpha ,\iota} v_{\alpha}
L_0^{\alpha ,\iota} \, ,
\end{equation}
whereas the momentum simply reads

\begin{eqnarray}
:\hat{P}: & = & \sum_{\alpha ,\iota} \iota L_0^{\alpha ,\iota} +
\sum_{\kappa ,\alpha ,\iota} \iota q_{F\alpha}
:\hat{N}_{\alpha,\iota}(\kappa):\nonumber\\
& = & \sum_{\kappa ,\alpha ,\iota}[\kappa + \iota q_{F\alpha}]
:\hat{N}_{\alpha,\iota}(\kappa):
+ {2\pi\over N_a}\sum_{\kappa ,\kappa'}
\sum_{\alpha ,\iota}{\iota\over 2}[
:\hat{N}_{\alpha,\iota}(\kappa)::\hat{N}_{\alpha,\iota}(\kappa '):
\nonumber\\
& + &
:\hat{N}_{\alpha,\iota}(\kappa)::\hat{N}_{\alpha,-\iota}(\kappa '):]
\, .
\end{eqnarray}
We emphasize that the absence of the interaction-dependent
parameters $\xi_{\alpha\alpha'}^j$ $(19)$ in
the quantity $\sum_{\alpha ,\iota} \iota L_0^{\alpha ,\iota}$
of the RHS of Eq. $(60)$ follows from the fact that
the matrix of entries $\xi_{\alpha\alpha'}^0$ is
the inverse of the transposition of the matrix
of entries $\xi_{\alpha\alpha'}^1$. Notice that the
two-pseudoparticle interaction terms of the normal-orderered
critical-point momentum operator $(60)$ are not present in the
original momentum expression (see Eq. $(22)$ of I).
This is because the latter expression refers to the
Hilbert space associated with a canonical ensemble
characterized by constant $(\eta_z,S_z)$ eigenvalues.
In turn, the two-pseudoparticle interaction terms of
the normal-orderered critical-point momentum operator
$(60)$ account for the changes in the pseudo-Fermi
points $(7)-(9)$ due to changes in the eigenvalues
$\eta_z$ and $S_z$: these interaction terms give
zero when acting on states with the same values
of $\eta_z$ and $S_z$ as the reference ground state.
We stress that both the Hamiltonian
expression $(59)$ and the momentum expression $(60)$
are valid {\it only} at the critical point. Therefore,
they lead to the correct excitation energy and
momentum, respectively, to second order in the
density of excited pseudoparticles {\it only}.

The HWS of the Virasoro algebras, $|h_{\alpha}^{\iota}\rangle$,
are eigenstates of the generators $(57)$ satisfying

\begin{equation}
L_0^{\alpha ,\iota}|h_{\alpha}^{\iota}\rangle =
h_{\alpha}^{\iota} |h_{\alpha}^{\iota}\rangle \, .
\end{equation}
The $\alpha ,s$ towers are constructed by applying the generators
$(56)$ to the HWS of the Virasoro algebras, {\it i.e.},

\begin{equation}
L^{\alpha ,\iota}_{-N^{\alpha\iota}_{ph}}
|h_{\alpha}^{\iota}\rangle =
(h_{\alpha}^{\iota} + N^{\alpha\iota}_{ph})|h_{\alpha}^{\iota}
+ N^{\alpha\iota}_{ph}\rangle \, ,
\end{equation}
where the states $|h_{\alpha}^{s} +
N^{\alpha\iota}_{ph}\rangle$ are also eigenstates of the
operators $L_0^{\alpha ,\iota}$

\begin{equation}
L_0^{\alpha ,\iota} |h_{\alpha}^{\iota} +
N^{\alpha\iota}_{ph}\rangle
= (h_{\alpha}^{\iota} + N^{\alpha\iota}_{ph})|h_{\alpha}^{\iota} +
N^{\alpha\iota}_{ph}\rangle \, .
\end{equation}

{}From Eqs. $(59)$ and $(60)$ we find that the energy spectrum
associated with all $\alpha ,\iota$ conformal families
is given by Eq. $(37)$ and that the corresponding momentum
is

\begin{equation}
P = \sum_{\alpha,\iota} \iota[h_{\alpha}^{\iota} +
N^{\alpha\iota}_{ph}] +
\sum_{\alpha} D_{\alpha} 2q_{F\alpha} \, .
\end{equation}

In the present class of quantum liquids the HWS of the
Virasoro algebras are the
Hamiltonian eigenstates (A) and (B) of form $(26),(27)$
and $(28),(29)$, respectively. There is a one-to-one
correspondence between the primary-fields operators and
the HWS \cite{Neto,Boy,Houches}. Therefore, the primary-field
operators correspond to
excitations which connect adiabatically ground states
$(25)$ of neighboring parameter-space points (excitations
(A)); and large-momentum, low-energy pseudoparticle-pseudohole
excitations (excitations (B)). In contrast to the states
(C), both the excitations (A) and (B) change the
conserved pseudoparticle numbers $N_{\alpha ,\iota}$.
The simple form of the generators of the states (A) and (B)
(see Eqs. $(26),(27)$
and $(28),(29)$, respectively) reveals that the corresponding
primary-field operators have simple expressions in
the pseudoparticle basis.

The interaction terms of the critical-point Hamiltonian
$(14),(24)$ are controlled by two-pseudoparticle forward
scattering through the corresponding pseudoparticle
phase shifts. These appear
in the excitation energies of the HWS of the Virasoro algebras
through the conformal dimensions $h^{\iota}_{\alpha}$
$(38)$ and, therefore, the interaction dependence of the
critical exponents associated with these HWS, as for example the
exponents of the oscillating terms in the correlation functions
\cite{Frahm}, is determined by the pseudoparticle
interactions \cite{Neto,Carmelo92}.

On the other hand, the energy spectrum of the small-momentum
and low-energy excitations (C),  which determines the
response of the energy-momentum tensor to the curvature of the
two-dimensional space-time, does {\it not} include, at the
critical point, the interaction-dependent
two-pseudoparticle phase shifts.
It follows that $c_{\alpha\iota}=1$ in Eq. $(54)$. This is
because the excitations (C) involve only the pseudoparticle
non-interacting Hamiltonian $(39)$ and correspond to the
towers $(62)$ whose spectrum is given, exclusively, in terms
of the integer numbers $(31)$.

\section{AFFINE LIE ALGEBRAS AND DYNAMICAL SEPARATION}

For each of the two (or $\nu $) above Virasoro algebras, there
is associated an affine-Lie algebra \cite{Neto}. These
affine-Lie algebras \cite{Halpern} are often
in the literature \cite{Neto,Boy,Houches}
called Kac-Moody algebras \cite{Kac,Moody}, although
the Kac-Moody algebras are in fact more general
and also include the hyperbolic algebras which are {\it not}
affine-Lie algebras \cite{Halpern}.

The affine-Lie algebras are generated by considering the conserved
quantities related to the original particles of the theory,
$N_{\gamma(\alpha)}=N_{\alpha}$, where $N_{\alpha}$
is the eigenvalue of the $\alpha$ - pseudoparticle
number operator introduced in I,
$\hat{N}_{\alpha} \equiv \sum_{q}\hat{N}_{\alpha }(q)
=\sum_{q} b^{\dag }_{q,\alpha}b_{q,\alpha}$,
and $\gamma(\alpha)$ is associated with the particle
conserving number, $N_{\gamma(\alpha)}$, which equals
the number of $\alpha$ pseudoparticles. For the Hubbard chain
(see I) $N_c=N=N_{\uparrow}+N_{\downarrow}$ and
$N_s=N_{\downarrow}$ and, therefore, $\gamma (c)=\rho$ (charge)
and $\gamma (s)=\downarrow$ (down spin). Since
$N_{\gamma(\alpha)}=N_{\alpha}$ and
$\hat{N}_{\gamma(\alpha)}=\hat{N}_{\alpha}$, it follows
that the particle-number operators $\hat{N}_{\gamma(\alpha)}$
can be written in the pseudoparticle basis as

\begin{equation}
\hat{N}_{\gamma(\alpha)} \equiv \sum_{q}\hat{N}_{\alpha }(q)
= \sum_{q} b^{\dag }_{q,\alpha}b_{q,\alpha}
\, .
\end{equation}

We can associate to each conserved quantity in $(65)$ a
$\gamma(\alpha)$ current. These two (or $\nu$) $\gamma(\alpha)$
currents are the diagonal generators of the corresponding
affine-Lie algebras \cite{Neto}. Using the notation of the
previous sections, we introduce the
following currents

\begin{eqnarray}
\hat{J}_0^{\gamma (\alpha )} =  \sum_{\iota}
\hat{N}_{\alpha ,\iota} \, ,\nonumber\\
\hat{J}_1^{\gamma (\alpha )} = \sum_{\iota} \iota
\hat{N}_{\alpha ,\iota} \, ,
\end{eqnarray}
which can be combined in the holomorphic and anti-holomorphic components,
$\hat{J}_{\iota }^{\gamma (\alpha )}= \hat{J}_0^{\gamma (\alpha )}+
\iota \hat{J}_1^{\gamma (\alpha )}$, such that

\begin{equation}
\hat{J}_{\iota}^{\gamma (\alpha )} =
2\hat{N}_{\alpha ,\iota} \, .
\end{equation}
(We denote $\hat{J}_{\iota }^{\gamma (\alpha )}$ with
$\iota=+1$ by $\hat{J}_{+1}^{\gamma (\alpha )}$, to
distinguish it from the current $\hat{J}_{1}^{\gamma (\alpha )}$
of Eq. $(66)$.)

Since the numbers $N_{\alpha ,\iota}$ are good quantum
numbers, at the critical point right (left)
pseudoparticles are made out of right (left)
electrons only. This implies that the particle currents
$\hat{J}_{\iota}^{\gamma (\alpha )}$ $(67)$ and
the pseudoparticle currents $\hat{J}_{\iota}^{\alpha }$
are equal. However, the associated particle current
densities, $\hat{J}_{\iota}^{\gamma (\alpha )}(x)$, and
pseudoparticle current densities, $\hat{J}_{\iota}^{\alpha }(x)$,
are different objects. This can be shown by considering the
Fourier transform of the current densities
$\hat{J}_{\iota}^{\gamma(\alpha)}(x)$ and
$\hat{J}_{\iota}^{\alpha }(x)$, which we call
$\hat{J}_{\iota}^{\gamma(\alpha)}(k)$
and $\hat{J}_{\iota}^{\alpha }(k)$, respectively.
As we show in Eqs. $(68)$ and $(69)$ below,
these two small-momentum currents are not
equal.

Following Ref. \cite{Neto}, the dressed-charge matrix \cite{Frahm}
is a representation of the small-momentum current
$\hat{J}_{\iota}^{\gamma(\alpha)}(k)$ in the reduced
$\iota$ Hilbert space of vanishing energy and momentum. This
$\iota$ space is spanned by the ground state $(25)$ and
two single-pair $\alpha$, $\iota$ - pseudoparticle
eigenstates (C) which are constructed from that ground
state by transferring the pseudoparticle at
$\kappa =0$ to $\kappa =\iota{2\pi\over {N_a}}$.
We call these excited eigenstates $|\alpha ,\iota;\eta_z,S_z\rangle$.
In Appendix A we use the methods introduced in
Ref. \cite{Carmelo92c} to show that at the smallest
momentum $k=\iota '{2\pi\over {N_a}}$ (which vanishes
as $N_a \rightarrow \infty$)

\begin{equation}
\langle\alpha ,\iota;\eta_z,S_z|\hat{J}_{\iota '}^{
\alpha '}(k)|0;\eta_z,S_z\rangle
= \delta_{\iota ,\iota'} \delta_{\alpha ,\alpha '}
\, ,
\end{equation}
whereas,

\begin{equation}
\langle\alpha ,\iota;\eta_z,S_z|\hat{J}_{\iota '}^{\gamma
(\alpha ')}(k)|0;\eta_z,S_z\rangle
= \delta_{\iota,\iota'} \xi_{\alpha \alpha '}^1
\, .
\end{equation}
Comparision of Eqs. $(68)$ and $(69)$ confirms that
although $\hat{J}_{\iota }^{\gamma (\alpha )}=\hat{J}_{\iota }^{
\alpha }$ we have that $\hat{J}_{\iota }^{\gamma (\alpha )}(k)\neq
\hat{J}_{\iota }^{\alpha }(k)$, and thus
$\hat{J}_{\iota }^{\gamma (\alpha )}(x)\neq
\hat{J}_{\iota }^{\alpha }(x)$.

{}From our previous discussions, it is clear that in
the phases of lowest symmetry and at low energies the
excitations of the system are pseudoparticle-pseudohole
excitations (B) and (C). Thus, the transport
carriers of the quantum liquid are the $\alpha$
pseudoparticles \cite{Carmelo92c}. These pseudoparticles
couple to external fields, whose couplings determine the exotic
instabilities observed in quasi-one-dimensional synthetic
metals \cite{Campbell}. The pure zero-momentum forward-scattering
character of the Hamiltonian $(1)$ implies that
for these energy scales the $\alpha$ pseudoparticle
currents are non-dissipative and contribute to
the coherent part of the $\gamma (\alpha)$ and other
conductivities only \cite{Neto,Carmelo92,Carmelo92c}.
(See the Lehmann representation of the general
$\vartheta $ conductivity spectrum in Eq.
(A10) of Appendix A.)

By introducing a magnetic flux through a ring (periodic
boundary conditions), it is possible to calculate the
charge stiffness associated with these pseudoparticles. This
is done by taking the second derivative of the energy with
respect to the flux \cite{Kohn}. On the other hand,
the charge stiffness can be shown to be proportional
to the weight of the $\delta$-peak which constitutes the
coherent part of the charge conductivity spectrum
\cite{Carmelo92,Carmelo92c,Shastry,Zotos}. (This is given
in the general case (charge, spin, spin projection) by
Eq. (A10) of Appendix A.) The stiffness determines the
transport masses of the pseudoparticles. These masses are defined
in Eq. $(59)$ of Ref. \cite{Carmelo92c} and are proportional
to the inverse of the pseudoparticle elementary currents
\cite{Carmelo92,Carmelo92c}.

Consider the real part of the frequency-dependent
conductivity, $\sigma^{\vartheta }(\omega )$, associated
with the current $\hat{J}_1^{\vartheta }$, Eq.
(A10) of Appendix A. (In the case of the Hubbard chain,
for example, $\vartheta$ can be charge $\vartheta=\rho$,
spin $\vartheta=\sigma_z$, and spin projection
$\vartheta=\sigma$.) In that Appendix we use the method
used in Ref. \cite{Carmelo92c} for charge $\vartheta=\rho$ and
spin $\vartheta=\sigma_z$ to evaluate the expression for
the general $\vartheta$ stiffness in terms of the
pseudoparticle transport masses. The result is

\begin{equation}
2\pi D^{\vartheta } = \sum_{\alpha '}
{q_{F\alpha '}\over {m_{\alpha '}^{\vartheta }}} \, ,
\end{equation}
where the transport masses are given by

\begin{equation}
m_{\alpha }^{\vartheta }\equiv
{q_{F\alpha }\over {[g^{\vartheta }_{\alpha}
\sum_{\alpha '}\sum_{\alpha ''} g^{\vartheta }_{\alpha ''} v_{\alpha
'}\xi^1_{\alpha '\alpha }\xi^1_{\alpha '\alpha ''}]}}.
\end{equation}
The latter equation represents the $\vartheta$-transport mass
for a $\alpha $, $\iota$ pseudoparticle of pseudomomentum
$\kappa =0$ ($q=\iota q_{F\alpha }$) \cite{Carmelo92c}.
In the RHS of Eqs. $(70)$ and $(71)$ the velocity
$v_{\alpha}$ and pseudomomentum $q_{F\alpha }$ are
given in Eqs. $(5)$ and $(10)$, respectively. The
interaction-dependent quantities $\xi^1_{\alpha\alpha '}$
are the elements of the dressed-charge matrix
$(19)$ which are the two-particle matrix elements
$(69)$ (when $\iota =\iota'$). The integer numbers
$g^{\vartheta }_{\alpha}$ are model dependent and are the
coefficients of the general equation (A3) of Appendix A. They
are given in that Appendix for the case of the Hubbard
chain.

The properties of the $\gamma (\alpha)$ currents, which are
the diagonal generators of the affine-Lie algebras associated
with the Virasoro algebras studied in Sec. IV, lead
to the concept of {\it dynamical separation} \cite{Neto}.
Indeed, consider $\vartheta = \gamma (\alpha )$ in the above
expressions. In Appendix A we find

\begin{equation}
2\pi D^{\gamma (\alpha )} = {q_{F\alpha }\over {m_{\alpha }^{\gamma
(\alpha) }}} \, ,
\end{equation}
where

\begin{equation}
m_{\alpha }^{\gamma (\alpha )} = {q_{F\alpha }\over
{[\sum_{\alpha '}v_{\alpha '}(\xi^1_{\alpha '\alpha })^2]}} \, ,
\end{equation}
and

\begin{equation}
m_{\alpha '}^{\gamma (\alpha) }=\infty \, ,
\end{equation}
for $\alpha \not= \alpha'$. Therefore, the stiffness of
the $\gamma (\alpha)$ conductivity spectrum involves
the transport mass (and elementary current) of the
corresponding $\alpha$ pseudoparticle {\it only}. This holds
true for the two (or $\nu$) $\gamma (\alpha)$ stiffnesses
only. It follows that in the present $U(1)\otimes U(1)$ sector
the peaks of the charge and down-spin conductivity
spectra are determined, in the Hubbard chain, by elementary currents
of $c$ and $s$  pseudoparticles only, respectively.

This does not hold true, however, for the non-$\gamma
(\alpha)$ conductivities whose stiffness expressions
involve a superposition of several elementary
currents of different $\alpha$ pseudoparticle
branches. This is the case of the spin and up-spin
conductivities of the Hubbard chain.

The concept of dynamical separation introduced in
\cite{Neto} associates the $\alpha$
pseudoparticles with the $\gamma (\alpha)$ conductivities.
It means that from the point of view of the two
$\gamma (\alpha)$ diagonal generators of the
affine-Lie algebras each $\alpha$ branch of
pseudoparticles can be identified with the
conserved quantity $\gamma (\alpha)$.
However, it will be shown elsewhere \cite{Two,Carmelo94a}
that from the point of view of the $\alpha$, $\iota$ generators
which transform the ground state $(25)$ onto the small-momentum and
low-energy one-pair $\alpha$, $\iota$ pseudoparticle excitations the
colors $\alpha$ {\it are not} the quantum numbers
$\gamma (\alpha)$. This follows from the properties of the
canonical transformation which at low energy connects the
electronic and pseudoparticle basis.

\section{CONCLUDING REMARKS}

Our results in the preceding sections establish that the
pseudoparticle basis and perturbation theory developed
in I can be applied to study the critical-point
physics of multicomponent integrable systems. Consistent
with earlier work based on the finite-size corrections
\cite{Izergin,Frahm}, we have found that the
complete critical theory of these systems is determined
by a direct product of Virasoro algebras. We were able
to write the expressions for the generators of these algebras
and other operators explicitly, {\it i.e.} we have introduced the
operator-Virasoro algebras of integrable quantum liquids.
Taken together, the results of I and the present paper
clarify considerably several points
concerning the non-Fermi liquid physics in one-dimensional
quantum liquids.

First, they have revealed that the correct critical-point
Hamiltonian of integrable systems refers to the
pseudoparticle basis.

Second, they have shown that despite the many-pseudoparticle
($k=0$ forward-scattering) interaction terms in the
critical-point Hamiltonian, the theory is
nonetheless conformal and scale invariant,
{\it i.e.} the pseudoparticle operator basis
permits a perturbative expansion around an interacting electron
fixed point.

Third, the Landau-liquid character of the $U(1)\otimes U(1)$
sectors of the Hubbard chain
\cite{Carmelo91,Carmelo91b,Carmelo92,Carmelo92b,Carmelo92c,Campbell}
follows from the fact that at low energy and at
small momentum the Hamiltonian pseudoparticle
interaction terms are {\it irrelevant}. Moreover, the validity
of the Landau-liquid expansion of Refs.
\cite{Carmelo91b,Carmelo92b,Carmelo92c}
was shown to follow from the perturbative character
of the quantum problem in the pseudoparticle operator
basis.

Fourth, our results reveal that the present kind of
many-body problem defined in a discrete lattice can be
mapped into a set of non-interacting quantum-field theories
at the critical point. However, despite
the correspondence between the pseudoparticle colors
$\alpha$ and the particle quantum numbers $\gamma (\alpha)$
in terms of the $\alpha$ elementary currents and
$\gamma (\alpha)$ conductivities (dynamical separation), we
will show elsewhere \cite{Two,Carmelo94a}
that, from the point of view of the generators of the low-energy and
small-momentum excitations, the pseudoparticle colors
$\alpha$ and the particle quantum numbers $\gamma (\alpha)$
{\it are not} equivalent. This means that the
quantum numbers $\alpha$ which label the above
set of independent and non-interacting quantum-field
theories {\it are not} usual quantum numbers such
as spin projection, charge, and spin. This also
implies the non-perturbative character of the usual
electronic basis, where the above quantum-field
theories are more difficult to describe.

Finally, it has been argued that non-Fermi liquid behavior occurs
in some sectors of parameter space of {\it two}-dimensional
interacting quantum liquids \cite{Ander,Varma,Guinea,Hua}.
If this proves true, some of the pseudoparticle concepts and techniques
developed in the present one-dimensional context
may prove useful in that case as well.

\nonum
\section{ACKNOWLEDGMENTS}

This work was supported by C.S.I.C. [Spain] and the University
of Illinois. We thank M. B. Halpern for constructive remarks on
affine-Lie algebras and Kac-Moody algebras, H. Frahm and V. E.
Korepin for useful comments concerning the expressions for the
dimensions of the fields, and F. H. L. Essler, E. H. Fradkin, F.
Guinea, P. Horsch, A. A. Ovchinnikov, N.M.R. Peres, and K. Maki for
stimulating discussions. The hospitality and support of the C.S.I.C.
and Department of Physics of the UIUC are gratefully acknowledged
by J. M. P. C.. A. H. C. N. thanks CNPq (Brazil) for a scholarship.
D.K.C. and A.H.C.N. acknowledge the partial support of this research
by the U.S. National Science Foundation under grants
NSF-DMR89-20538 and NSF-DMR91-22385, respectively.

\vfill
\eject
\appendix{CURRENT REPRESENTATION AND\\
          COHERENT PART OF THE CONDUCTIVITY SPECTRA}

In this Appendix we use the methods used in Ref. \cite{Carmelo92c}
for $\vartheta=\rho$ and $\vartheta=\sigma_z$, to derive the
current representations of Eqs. $(68)$ and $(69)$ and the stiffness
expressions $(70)-(74)$ for general particle quantum
numbers $\vartheta$. Although we consider mainly the
case of the Hubbard chain, our results and expressions
are valid for multicomponent integrable quantum liquids.

Equation $(68)$ follows from the fact that
$\hat{J}_{\iota }^{\alpha }(k)|0;\eta_z,S_z\rangle
=|\alpha ,\iota;\eta_z,S_z\rangle$.

Let us consider the $\vartheta -\vartheta$ correlation
function, $\chi^{\vartheta }(k,\omega)$. (In
the case of the Hubbard chain $\vartheta$ can denote
charge ($\vartheta=\rho$), spin ($\vartheta=\sigma_z$),
and spin projection ($\vartheta=\sigma$).)

At the smallest momentum $k=\iota{2\pi\over {N_a}}$
(which vanishes as $N_a \rightarrow \infty $ and
vanishing frequency $\omega$, the only excited eigenstates
are the single-pair pseudoparticle states $|\alpha
,\iota;\eta_z,S_z\rangle$ of Eq. $(69)$. Therefore, the
use of a Lehmann representation for $\chi^{\vartheta
}(k,\omega)$ gives when $k=\iota{2\pi\over {N_a}}$ and
$\omega $ is vanishing small \cite{Carmelo92c}

\begin{equation}
\chi^{\vartheta }(k,\omega) = -
{N_a\over {\pi}}
\sum_{\alpha}|\langle\alpha ,\iota;\eta_z,S_z|
\hat{J}_{\iota }^{\vartheta }(k)
|0;\eta_z,S_z\rangle|^2 {k^2v_{\alpha}\over {(kv_{\alpha})^2 -
(\omega+i\eta)^2}} \, .
\end{equation}

In order to derive the matrix elements $\langle\alpha
,\iota;\eta_z,S_z|\hat{J}_{\iota }^{\vartheta }(k)
|0;\eta_z,S_z\rangle$ in (A1) we evaluate
$\chi^{\vartheta }(k,\omega)$ by a second method.
Comparison of the obtained expression with
(A1) and the use of a boundary condition
leads then to Eq. $(69)$.

Let us couple the pseudoparticles to a weak external
$\vartheta$ probe of long wavelength and low frequency.
According to the fluctuation-dissipation theorem,
the correlation function $\chi^{\vartheta }(k,\omega)$
equals the response function

\begin{equation}
\chi^{\vartheta }(k,\omega) =
{\delta \langle \hat{N}_{\vartheta }(k,\omega)\rangle
\over {V^{\vartheta }(k,\omega)}}  \, ,
\end{equation}
where $\hat{N}_{\vartheta }$ is the $\vartheta$ particle
conserving number, $e^{\vartheta }V^{\vartheta }(k,\omega)$
is the external potential, and $e^{\vartheta }$
is the corresponding elementary constant. For
instance, $e^{\rho }=-e$ is the electronic charge,
$e^{\sigma_z }=1/2$ is the electronic spin, and
$e^{\sigma }=\sigma 1/2=\pm 1/2$ are the
electronic spin projections.

The particle number $N_{\vartheta }$ can be written in terms
of the numbers $N_{\gamma (\alpha)}=N_{\alpha}$ as follows

\begin{equation}
N_{\vartheta} = \sum_{\alpha} g^{\vartheta }_{\alpha }
N_{\alpha} \, ,
\end{equation}
where $g^{\vartheta }_{\alpha }$ are model-dependent
integers. In the case of the Hubbard chain, where
$\gamma (c)=\rho$ and $\gamma (s)=\downarrow$, they
read $g^{\rho }_{c}=1$, $g^{\rho }_{s}=0$,
$g^{\downarrow }_{c}=0$, $g^{\downarrow }_{s}=1$,
$g^{\uparrow }_{c}=1$, $g^{\uparrow }_{s}=-1$,
$g^{\sigma_z }_{c}=1$, and $g^{\sigma_z }_{s}=-2$.
The following result holds true for all
multicomponent models :

\begin{equation}
g^{\gamma (\alpha)}_{\alpha '} = \delta_{\alpha ,\alpha '}
\, .
\end{equation}
As discussed in Ref. \cite{Carmelo92c}, the quantity

\begin{equation}
C^{\vartheta }_{\alpha } = e^{\vartheta}
g^{\vartheta }_{\alpha } \, ,
\end{equation}
is the coupling of the $\alpha $ pseudoparticles to
the $\vartheta $ probe.

The suitable inhomogeneous fluctuation $\delta \langle
\hat{N}_{\vartheta }(k,\omega)\rangle$ to be used
in Eq. (A2) is of the form \cite{Carmelo92c}

\begin{equation}
\delta \langle \hat{N}_{\vartheta }(k,\omega)\rangle
= {N_a\over {2\pi }}\sum_{\alpha}
\int_{q_{\alpha}^{(-)}}^{q_{\alpha}^{(+)}}
\delta N_{\alpha}(q;k,\omega ) g^{\vartheta }_{\alpha }
\, ,
\end{equation}
where the two inhomogeneous pseudomomentum deviations
$\delta N_{\alpha}(q;k,\omega )$ (with $\alpha = c,s)$
are determined by the following system of two
(or $\nu$) kinetic equations

\begin{eqnarray}
(kv_{\alpha}(q)-\omega)\delta N_{\alpha}(q;k,\omega)
+(sgn(q))k\delta(q_{F\alpha}-|q|)\sum_{\alpha '}
{1\over 2\pi}\int_{q_{\alpha '}^{(-)}}^{q_{\alpha '}^{(+)}}dq'
\delta N_{\alpha '} (q';k,\omega)
f_{\alpha\alpha '}(q,q') & + &\nonumber\\
g^{\vartheta }_{\alpha } V^{\vartheta }(k,\omega) = 0 \, ,
\hspace{1cm} \alpha = c, \, s \, .
\end{eqnarray}
In order to evaluate the response function (A2) we have to
solve the coupled kinetic equations (A7). The form of the
general solutions for these equations is

\begin{equation}
\delta N_{\alpha}(q;k,\omega) = - {1\over {N_a}}\left[
\delta(q_{F\alpha}-|q|){sgn(q)k\over {kv_{\alpha}(q)-\omega-i\eta}}
X_{\alpha}^{\vartheta }(q;k,\omega)\right] \, ,
\end{equation}
where the imaginary term, $i\eta$, conforms to the usual
adiabatic boundary conditions. The equations which determine
the functions $X_{\alpha}^{\vartheta }(q;k,\omega)$ are
obtained by replacing the  expressions (A8) in the
kinetic equations. Following Eqs. (A2), (A6), and (A8)
the response function is given by

\begin{equation}
\chi^{\vartheta }(k,\omega) =
-{N_a\over {2\pi V^{\vartheta }(k,\omega)}}
\sum_{j=\pm 1}\sum_{\alpha}
{k \over {kv_{\alpha}-j(\omega+i\eta)}}
g^{\vartheta }_{\alpha }
X_{\alpha}^{\vartheta }(jq_{F\alpha};k,\omega) \, .
\end{equation}
Evaluation of the functions $X_{\alpha}^{\vartheta
}(jq_{F\alpha};k,\omega)$ \cite{Carmelo92c} leads to
expression (A1) with the amplitudes $|\langle\alpha
,\iota;\eta_z,S_z|\hat{J}_{\iota }^{\vartheta }(k)
|0;\eta_z,S_z\rangle|$ expressed in terms of the
$j=1$ phase-shift combinations $(19)$. Finally,
the use of a suitable boundary conditions provides
the relative phases of the matrix elements with
the result

\begin{equation}
\langle\alpha ,\iota;\eta_z,S_z|\hat{J}_{\iota }^{\vartheta }(k)
|0;\eta_z,S_z\rangle = \sum_{\alpha '}g_{\alpha '}^{\vartheta }
\langle\alpha ,\iota;\eta_z,S_z|\hat{J}_{\iota }^{\gamma
(\alpha ')}(k)|0;\eta_z,S_z\rangle \, ,
\end{equation}
where $\langle\alpha ,\iota;\eta_z,S_z|\hat{J}_{\iota }^{\gamma
(\alpha ')}(k)|0;\eta_z,S_z\rangle$ is
given by Eq. $(69)$ with $\iota '=\iota$. (It is
obvious that $(69)$ vanishes when $\iota '\neq \iota$.)

In order to calculate the $\vartheta$ stiffnesses
$(70)-(74)$ we use the correlation-function expression
(A1) with the matrix elements given by Eq.
$(69)$. The Lehmann representation of the $\vartheta$
conductivity spectrum reads \cite{Carmelo92c}

\begin{equation}
Re \, \sigma^{\vartheta }(\omega) =
2\pi (e^{\vartheta })^2 \left[ D^{\vartheta }\delta (\omega) +
{1\over {N_a}}\sum_{i\neq 0}
|\langle i|\hat{J}_1^{\vartheta }|0;\eta_z,S_z \rangle |^2
\delta \left(\omega_{i,0}^2-\omega ^2\right) \right] \, ,
\end{equation}
where the $i$ summation refers to both the ``dissipative''
LWS II and corresponding non-LWS multiplet eigenstates
of zero momentum, which we denote by $|i \rangle$.
In the lowest-symmetry sectors of parameter space they have
{\it finite} excitation energy $\omega_{i,0}=E_i-E_0$
($E_0$ denotes the ground-state energy). $D^{\vartheta }$ is
the $\vartheta$ stiffness. In order to derive $D^{\vartheta }$,
we use the following standard relation of the $\vartheta$
conductivity spectrum (A11) to the response function
$\chi^{(\vartheta)}(k,\omega)$ \cite{Carmelo92c}

\begin{equation}
Re \, \sigma^{\vartheta } (\omega) =
Re \left[ \lim_{k \to 0}
i(e^{\vartheta })^2{\omega\over {k^2}}{\chi^{\vartheta }(k,\omega)
\over {N_a }} \right] \, .
\end{equation}
In the present lattice model this relation
is only valid for $\omega\rightarrow 0$ and at
$\omega = 0$. Therefore, it provides the coherent
part of (A11). The use of (A1) and $(69)$ in
the RHS of Eq. (A12) leads to

\begin{equation}
Re \, \sigma^{\vartheta } (\omega) =
2\pi (e^{\vartheta })^2 D^{\vartheta }\delta (\omega)
\, ,
\end{equation}
with $D^{\vartheta }$ given by Eqs. $(70)-(71)$.
The $D^{\gamma (\alpha )}$ expression $(72)-(74)$
follows from Eqs. $(70)-(71)$ and (A4).


$*$ Permanent address,
Department of Physics, University of \'Evora, Apartado 94, P-7001
\'Evora CODEX, Portugal.


\begin{references}
\bibitem[1]{Carmelo94}
        J. M. P. Carmelo, A. H. Castro Neto, and
        D. K. Campbell, preprint UIUC (1994).
\bibitem[2]{Yang}
        For one of the first generalizations of the Bethe
        ansatz to multicomponent $(\nu=2)$ systems see
        C. N. Yang, Phys. Rev. Lett. {\bf 19}, 1312
        (1967).
\bibitem[3]{Lieb}
        Elliott H. Lieb and F. Y. Wu, Phys. Rev. Lett. {\bf 20},
        1445 (1968).
\bibitem[4]{Izergin}
        A.G. Izergin, V.E. Korepin, and N. Yu Reshetikhin,
        J. Phys. A: Math. Gen. {\bf 22}, 2615 (1989).
\bibitem[5]{Frahm}
        Holger Frahm and V. E. Korepin, Phys. Rev. B {\bf 42},
        10553 (1990); {\bf 43}, 5653 (1991).
\bibitem[6]{Neto}
        J. M. P. Carmelo and A. H. Castro Neto,
        Phys. Rev. Lett. {\bf 70}, 1904 (1993).
\bibitem[7]{Belavin}
        A. A. Belavin, A. M. Polyakov, and A. B
        Zamolodchikov, J. Stat. Phys. {\bf 34},
        763 (1984); Nucl. Phys. {\bf B241}, 333 (1984).
\bibitem[8]{Blote}
        H.W.J. Bl\"ote, John L. Cardy, and M.P. Nightingale,
        Phys. Rev. Lett. {\bf 56}, 742 (1985);
        Ian Affleck, Phys. Rev. Lett. {\bf 56}, 746 (1985).
\bibitem[9]{Boy}
        D. Boyanovsky and C.M. Naon, Rev. Nuo. Cim. {\bf 13},
        N. 2 (1990).
\bibitem[10]{Houches}
        For an important collection of relevant articles, see,
        {\it Fields, Strings and Critical Phenomena},
        (Les Houches 1988) edited by E. Br\'ezin and
        J. Zinn-Justin (North-Holland, 1990).
\bibitem[11]{Korepin}
        Fabian H. L. Essler, Vladimir E. Korepin,
        and Kareljan Schoutens, Phys. Rev. Lett. {\bf 67},
        3848 (1991); Nucl. Phys. B {\bf 372}, 559 (1992),
        and references therein.
\bibitem[12]{Woy}
        F. Woynarovich, J. Phys. A {\bf 22}, 4243 (1989).
\bibitem[13]{Pines}
        D. Pines and P. Nozi\`eres, in {\em The Theory of
        Quantum Liquids},
        (Addison-Wesley, Redwood City, 1966 and 1989), Vol. I.
\bibitem[14]{Baym}
        Gordon Baym and Christopher J. Pethick, in
        {\em Landau Fermi-Liquid Theory Concepts and Applications},
        (John Wiley \& Sons, New York, 1991).
\bibitem[15]{Carmelo91}
        J. Carmelo and A. A. Ovchinnikov, J. Phys.: Condens.
        Matter {\bf 3}, 757 (1991).
\bibitem[16]{Carmelo91b}
        J. Carmelo, P. Horsch, P.-A. Bares, and A. A. Ovchinnikov,
        Phys. Rev. B {\bf 44}, 9967 (1991).
\bibitem[17]{Carmelo92}
        J. M. P. Carmelo and P. Horsch,
        Phys. Rev. Lett. {\bf 68}, 871 (1992).
\bibitem[18]{Carmelo92b}
        J. M. P. Carmelo, P. Horsch, and A. A. Ovchinnikov,
        Phys. Rev. B {\bf 45}, 7899 (1992).
\bibitem[19]{Carmelo92c}
        J. M. P. Carmelo, P. Horsch, and A. A. Ovchinnikov,
        Phys. Rev. B {\bf 46}, 14 728 (1992).
\bibitem[20]{Campbell}
        J. M. P. Carmelo, P. Horsch, D. K. Campbell, and
        A. H. Castro Neto, Phys. Rev. B {\bf 48}, 4200
        (1993).
\bibitem[21]{Castro}
        Walter Metzner and Carlo Di Castro, Phys. Rev. B
        {\bf 47}, 16107, (1993).
\bibitem[22]{Two}
        J. M. P. Carmelo, D. K. Campbell, A. H. Castro Neto,
        and N. M. R. Peres, preprint UIUC (1994).
\bibitem[23]{Carmelo94a}
        J. M. P. Carmelo, A. H. Castro Neto, D. K. Campbell,
        and N. M. R. Peres, preprint UIUC (1994).
\bibitem[24]{Nuno}
        J. M. P. Carmelo and N. M. R. Peres, preprint CSIC (1994).
\bibitem[25]{Ander}
        Philip W. Anderson, Phys. Rev. Lett. {\bf 64},
        1839 (1990); {\bf 65} 2306 (1990);
        P. W. Anderson and Y. Ren, in {\it High Temperature
        Superconductivity}, edited by K. S. Bedell,
        D. E. Meltzer, D. Pines, and J. R. Schrieffer
        (Addison-Wesley, Reading, MA, 1990).
\bibitem[26]{Halpern}
        K. Bardak\c{c}i and M. B. Halpern, Phys. Rev. D {\bf 3},
        2493 (1971); M. B. Halpern and N. A. Obers preprint
        (1992).
\bibitem[27]{Kac}
        V. G. Kac, Funct. Anal. App. {\bf 1}, 328 (1967).
\bibitem[28]{Moody}
        R. V. Moody, Bull. Am. Math. Soc. {\bf 73}, 217 (1967).
\bibitem[29]{Kohn}
	W. Kohn, Phys.Rev. {\bf 133}, A171, (1964).
\bibitem[30]{Shastry}
        B. Sriram Shastry and Bill Sutherland,
        Phys. Rev. Lett. {\bf 65}, 243 (1990).
\bibitem[31]{Zotos}
        X. Zotos, P. Prelovsek, and I. Sega,
        Phys. Rev. B {\bf 42},
        8445 (1990).
\bibitem[32]{Varma}
        C. M. Varma, P. B. Littlewood, S. Schmitt-Rink,
        E. Abrahams, and A. E. Ruckenstein,
        Phys. Rev. Lett. {\bf 63}, 1996 (1989).
\bibitem[33]{Guinea}
        F. Guinea, E. Louis, and J. A. Verg\'es,
        Phys. Rev. B {\bf 45}, 4752 (1992).
\bibitem[34]{Hua}
        Hua Chen and Daniel Mattis, preprint University of
        Utah (1992).
\end{references}
\end{document}